\documentclass[11pt]{article}

\usepackage{amssymb}
\usepackage{graphicx}

\usepackage{authblk}

\usepackage[latexsym]{}
\usepackage{amssymb}
\usepackage{amsmath}
\usepackage{graphicx}
\usepackage{epsfig}
\usepackage{url}
\usepackage{algorithm}
\usepackage{algorithmic}
\usepackage{stmaryrd}
\usepackage{qtree}

\usepackage{tikz} 
\usetikzlibrary{positioning} 

\usepackage{rotating} 

\def\lojoin{\hbox{\raise -.2em\hbox to-.32em{$\urcorner$} \hbox to-.08em{$\lrcorner$} $\Join$}\,}

\usepackage{url}

\begin{document}
\title{An introduction to Graph Data Management\thanks{Work funded by the Millennium Nucleus Center for Semantic Web Research under Grant NC120004.}}
\author[1,3]{Renzo Angles}
\author[2,3]{Claudio Gutierrez}
\affil[1]{Dept. of Computer Science, Universidad de Talca, Chile (rangles@utalca.cl)}
\affil[2]{Dept. of Computer Science, Universidad de Chile, Chile (cgutierr@dcc.uchile.cl)}
\affil[3]{Center for Semantic Web Research (http://ciws.cl/)} 
\date{}

\maketitle

\begin{abstract}
A graph database is a database where the data structures for the
schema and/or instances are modeled as a (labeled)(directed) graph or
generalizations of it,  and where querying is expressed by
graph-oriented operations and type constructors. 
 In this article we present the basic notions of graph databases, 
give an historical overview of its main development, and study the
main current systems that implement them.
\end{abstract}



\section{Introduction}

 It has been long recognized that graphs are a natural way to
 represent information and knowledge. 
 In fact, the notion of ``graph database'' 
has a long development, at least since the 1980's.
But it is only recently that several technological
developments have made it possible to make this abstract idea a
reality. 
 Powerful hardware to store and process graphs, 
 powerful sensors to record directly the information,
 powerful machines that allow to analyze and visualize graphs, 
 among other factors, have given rise to the current flourishing 
 in the area of graph data management.

There are two broad and interrelated topics in this area that from our 
perspective deserve to be treated separately today. 
One is the area of graph database models, which comprises general principles 
that ideally should guide the design of systems.
The second is graph database systems themselves, which are system that deal with
graph data processing, sometimes addressing directly demands of users,
thus emphasizing factors such as efficiency, usability and direct
solutions to urgent data management problems.

\paragraph{Graph database models.}
The fundamental abstraction behind a database system is its database model. 
In the most general sense, a database model (or just data model) is a conceptual tool used to model representations of real world entities and the relationships among them.
 As is well known, a data model can be characterized by three basic components, namely data structures, query and transformation language, and integrity constraints. 
Following this definition, a graph database model is a model where data structures for the schema and/or
instances are modeled as graphs (or generalizations of them), 
where the data manipulation is expressed by graph-oriented operations (i.e. a graph query language), 
and appropriate integrity constraints can be defined over the graph structure

The main characteristic of a graph database is that the data are
conceptually modeled and presented to the user as a graph, that is,
the {\em data structures} (data and/or schema) are represented by graphs, or by data structures generalizing the notion of graph (e.g. hypergraphs or hypernodes). 
One of the main features of a graph structure is the simplicity to model unstructured data. Therefore, in graph models the separation between schema and data (instances) is less marked than in the classical relational model.

Regarding {\em data manipulation and querying}, it is is expressed by graph
transformations,  or by operations whose main primitives are based on graph features like paths, neighborhoods, subgraphs, graph patterns, connectivity, and graph statistics (diameter, centrality, etc.).
 Some graph models define a flexible collection of type constructors and
 operations, which are used to create and access the graph data
 structures.  Another approach is to express all queries using a few
 powerful graph manipulation primitives. Usually the query language is
 what gives a database model its particular flavor. In fact, the
 differences among graph data structures are usually minors as
 compared to   differences among graph query languages.

 Finally, {\em integrity constraints} enforce data consistency. 
 These constraints can be grouped in schema-instance consistency, identity and referential integrity, and functional and inclusion dependencies.
 Examples of these are  labels with unique names,
 typing constraints on nodes  functional dependencies,
 domain and range of properties, etc.

 In this article we will concentrate in the data structure and language
facets of graph database models.

\paragraph{Graph Data Management Systems}
There are two categories of graph data management systems: graph databases and graph processing frameworks. 
 These systems provide two perspectives for storing and querying graph data, each one with their own goals.
 The former are systems specifically designed for managing graph-like
data following the basic principles of database systems, i.e. persistent data storage, physical/logical data independence, data integrity and consistency. 
 The latter are frameworks for batch processing and analysis of big graphs putting emphasis in the use of multiple machines to improve the performance 
 

\paragraph{Contents and Organization of this article.}
This article presents an overview of the basic notions, the 
historical evolution and the main current developments of this area.
There are three main topics, distributed by sections. 
First, an overview of the field and its development, which we hope
can be of help to look for ideas and past experiences.
Second, a review of the main graph database models in order to give
a perspective on actual developments.
Third, a similar review of graph database query languages.
Finally, we present current graph data management systems in a comparative manner.


\section{Motivation and Overview of the Field}

In this section we present motivations for graph data management
and briefly review the developments of it.
There is an emphasis on models in order to give a certain
abstraction level and unity of concepts that sometimes get lost
in the wide diversity syntaxes and implementation
solutions that exist today.
This section follows closely our review \cite{90004}.

\subsection{Why graph database models?}
 The first question one should answer is 
why to choose a graph data model instead of a relational,
 object-oriented,  semi-structured, or other type of data model.
The one-sentence answer is: 
Graph models are designed to manage data  in areas where the
main concern has to do with the interconnectivity
or topology of that data. 
In these applications, the atomic data and the relations amongst
the units of data have the same level of importance.

Among the main advantages that graph data models offer over other types of models, we can mention:

\begin{itemize}
\item Graphs has been long ago recognized as one of the most simple, natural and
  intuitive knowledge representation systems.
  This simplicity  overcomes  the limitations of the lineal format of classical writing systems. 

\item Graph data structures allow for a natural modeling when data has graph structure.
Graphs have the advantage of being able to keep all the information
about an entity in a single node and show related information by arcs connected to it.
Graph objects (like paths, neighborhoods) may have first order citizenship.

\item Queries can address direct and explicitly this graph structure.
Associated with graphs are specific graph operations in the query language algebra, such as finding shortest paths, determining certain subgraphs, and so forth. 
Explicit graphs and graph operations allow users to express a query at a high level of abstraction. In summary, Graph models realize for graph data the separation of concerns between modeling (the logic level) 
and implementation (physical level).

\item Implementation-wise, graph databases may provide special graph storage 
structures, and take advantage of efficient graph algorithms available 
for implementing specific graph operations over the data.
\end{itemize}

\subsection{Comparison with classical models} 
\label{sec:comparison}

As is well known, there are manifold approaches to model information and 
knowledge, depending on application areas and user needs.
We will briefly review the most influential of those models (relational,
semantic, object-oriented, semistructured)  and compare
them to graph data models.

The {\em Relational data model}~\cite{50228} was introduced by
Codd and is based on the simple notion of relation, which together with its associated algebra and logic, made the relational model a primary model for database research. 
In particular, its standard  query and transformation language, SQL, became a paradigmatic language for querying. 
It popularized the concept of abstraction levels by introducing a
separation between the physical and logical levels. 
Gradually the focus shifted to modeling data as seen by applications
and users (that is, tables). 
The differences between graph data models and the relational data model are manifold. 
The relational model is geared towards simple record-type data, where the data structure is known in advance (airline reservations, accounting, inventories, etc.). 
The schema is fixed, which makes it difficult to extend these databases. 
It is not easy to integrate different schemas, nor is it automatized. 
 The table-oriented abstraction is not suitable to naturally explore the underlying graph of relationships among the data, such as paths, neighborhoods, patterns.

{\em Semantic data models}~\cite{50200} focus on the incorporation
of richer and more expressive semantics into the database, from a user's viewpoint. 
Database designers can represent objects and their relations in a natural and clear manner (similar to the way users view an application) 
by using high-level abstraction concepts such as aggregation, classification and instantiation, sub- and super-classing, attribute inheritance and hierarchies.
A well-known and successful case is the entity-relationship model~\cite{50240}, which has become a basis for the early stages of database design.
Semantic data models are relevant to graph data model research because the semantic data models reason about the graph-like structure generated by the relationships between the modeled entities.

{\em Object-oriented (O-O) data models}~\cite{50190} are designed to
address the weaknesses of  the relational model in data intensive domains involving complex data objects and complex object interactions, such as CAD/CAM software, computer graphics and information retrieval.
According to the O-O programming paradigm on which these models are
based, they represent data as a collection of objects that are
organized into classes, and have complex values and methods. 
O-O data models are related to graph data models in their explicit or implicit use of graph structures in definitions.
Nevertheless, there are important differences with respect to the approach for modeling how to model the world. 
O-O data models view the world as a set of complex objects having certain state (data), where interaction is via method passing. 
On the other hand, graph data models view the world as a network of relations, emphasizing data interconnection, and the properties of these relations.
O-O data models focus on object dynamics, their values and methods.
Graph data models focus instead on the interconnection, while maintaining the structural and semantic complexity of the data.

{\em Semistructured data models}~\cite{50103,50122} were motivated by the increased existence of semistructured data (also called unstructured data), data exchange, and data browsing mainly on the Web.
In semistructured data, the structure is irregular, implicit and partial; the schema does not restrict the data, it only describes it, a feature that allows extensible data exchanges; the schema is large and constantly evolving; the data is self-describing, as it contains schema information.
Representative semistructured models are OEM~\cite{50155} and Lorel~\cite{50154}.
Many of these ideas can be seen in current
semi-structured languages like XML or JSON.
Generally, semistructured data is represented using a tree-like structure. 
However, cycles between data nodes are possible, which leads to graph-like structures like in graph data models.
Some authors characterize semistructured data as rooted directed connected graphs.

\subsection{Historical overview}
\label{historical}
 The ideas of graph databases can be dated at least to the nineties, 
 where much of the theory developed. 
Probably due to the lack of hardware support to manage big graphs,
this line of research declined for a while until a few years ago,
when processing graphs became common and 
a second wave of research was initiated.

\paragraph{The first wave.}
  In an early approach, facing the failure of
  contemporary systems to take into account the semantics of a
  database, a semantic network to store data about the database was
  proposed by Roussopoulos and Mylopoulos \cite{50163} .
 An implicit structure of graphs for the data itself was presented in the Functional Data Model~\cite{50184}, whose goal was to provide
a ``conceptually natural'' database interface.
 A different approach proposed the Logical Data Model (LDM)~\cite{50173}, where an explicit graph data model intended to generalize the relational, hierarchical and network models.
 Later, Kunii \cite{50165} proposed a graph data model for representing complex structures of knowledge called G-Base.

 In the late eighties an object-oriented data model based on a graph structure, called
O$_{2}$, was introduced by L\'ecluse et al \cite{50175}.
 Along the same lines, GOOD~\cite{50073} is an influential graph-oriented object model, intended to be a theoretical basis for a system in which manipulation as well as representation are transparently graph-based. 
 Among the subsequent developments based on GOOD are: 
 GMOD~\cite{50114} that proposes a number of concepts for graph-oriented database user interfaces;
 Gram~\cite{50019} which is an explicit graph data model for hypertext data;
 PaMaL~\cite{50269} which extends GOOD with explicit representation of tuples and sets;
 GOAL~\cite{50082} that introduces the notion of association nodes; 
 G-Log~\cite{50081} which proposed a declarative query language for graphs;
 and GDM~\cite{50298} that incorporates representation of n-ary symmetric relationships.

 There were proposals that used generalization of graphs with data modeling purposes.
The Hypermodel \cite{50175} (which we will develop in more detail)
was a model based on nested graphs on which subsequent work was developed \cite{50168,50152}.
 The same idea was used for modeling multi-scaled networks~\cite{50403} and genome data~\cite{50462}.

Another generalization of graphs, hypergraphs, gave rise to another
family of models.
 GROOVY~\cite{50187} is an object-oriented data model based on hypergraphs. 
 This generalization was used in other contexts:
 query and visualization in the Hy+ system~\cite{50189};
 modeling of data instances and access to them~\cite{50188};
 representation of user state and browsing~\cite{50185};

 There are several other proposals that deal with graph data models.
 G\"uting proposed GraphDB~\cite{50047} intended for modeling and querying graphs in object-oriented databases and motivated by managing information in transport networks.
 Database Graph Views~\cite{50145} proposed an abstraction mechanism to define and manipulate graphs stored in either relational object-oriented or file systems. 
 The project GRAS~\cite{50159} uses attributed graphs for modeling complex information from software engineering projects.
 The well known OEM~\cite{50155} model aims at providing integrated access to heterogeneous information sources, focusing on information exchange. 

Another important line of development has to do with data
representation models  and the World Wide Web. 
 Among them are data exchange models like XML~\cite{20004}, metadata representation models like RDF~\cite{10008} and ontology representation models like OWL~\cite{10129}. 

\paragraph{The second wave.}
We are witnessing the second impulse of development of graph data
management  which is focused on one hand, in practical systems, and on
the other, 
in theoretical analyses particularly of graph query languages. 
We will review the
former in Section \ref{systems} concentrating in database systems 
and will leave the latter out of this article.
 The reader interested in graph query languages can review article ``Foundations of Modern Query Languages for Graph Databases''\cite{survey2017}.

\section{Graph Database Models}

All graph data models have as their formal foundation variations on the
basic mathematical definition of a graph, e.g., directed or undirected
graphs, labeled or unlabeled edges and nodes, properties on nodes and edges, hypergraphs, hypernodes.

The most simple model is a plain labeled graph, ie. a graph with  nodes
and edges as everyone knows it.
 Although highly easy to learn, it has the drawback that it is difficult to modularize the information it represents. 
The notions of hypernodes and hypergraphs address this problem.
Hypergraphs, by enhancing the notion of simple edge,  allow the
representation of multiple complex relations.
On the other hand, hypernodes modularize the notion of node, by
allowing nesting graphs inside nodes.
 As drawbacks, both models use complex data structures which make it less intuitive their use and implementation.  

 Regarding simplicity, one of the most popularized models is the semistructured model, which use the most simple version of a graph, namely a tree, the most common and intuitive way or  organizing our data (e.g. directories)
 Finally, the most common models are slightly enhanced version of the
plain graphs. One of them, the RDF model, gives a light typing to nodes, and
considers edges as nodes, giving uniformity to the information objects
in the model.
 The other, the property graph model,  allows to adds properties to edges and nodes.

\begin{figure}[t!]
 \centering
 \includegraphics[width=12cm]{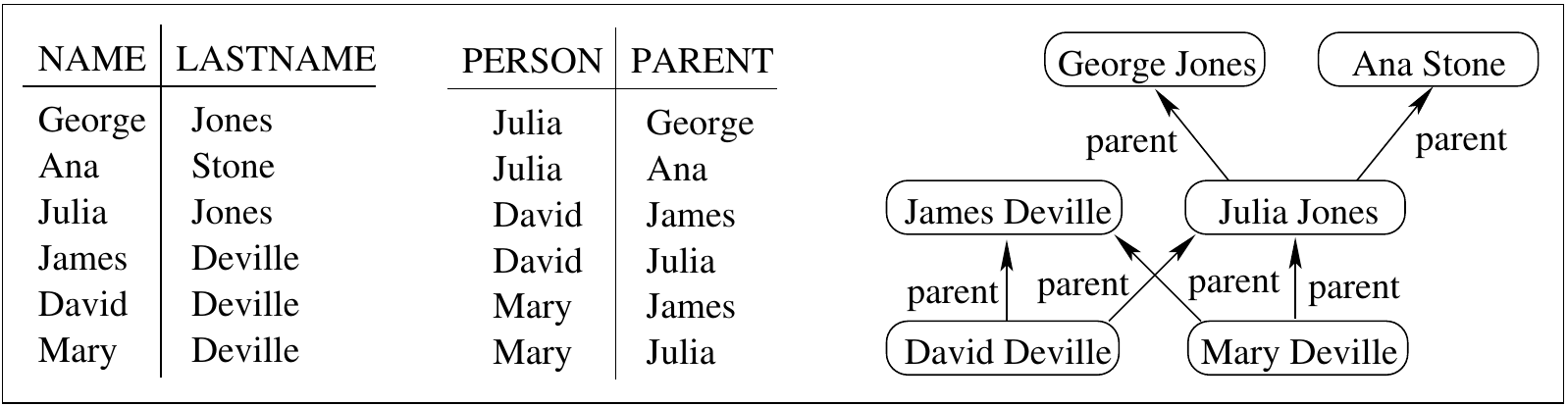}
 \caption{Example of a genealogy expressed in the relational model (i.e. as tables on the left) and a diagram of its scheme on the right.}
 \label{fig:basic}
\end{figure}

Next, we will present  these models and   show a
paradigmatic example of each. We will use the genealogy toy example
modeled as tables and a simple schema in Figure \ref{fig:basic}.

\subsection{The basics: Labeled graphs}
The most basic data structure for graph database models is a directed
graph with nodes and edges labeled by some vocabulary.
 A good example is Gram \cite{50019}, a graph data model motivated by
hypertext querying.

\begin{figure}[t!]
 \centering
 \includegraphics[width=12cm]{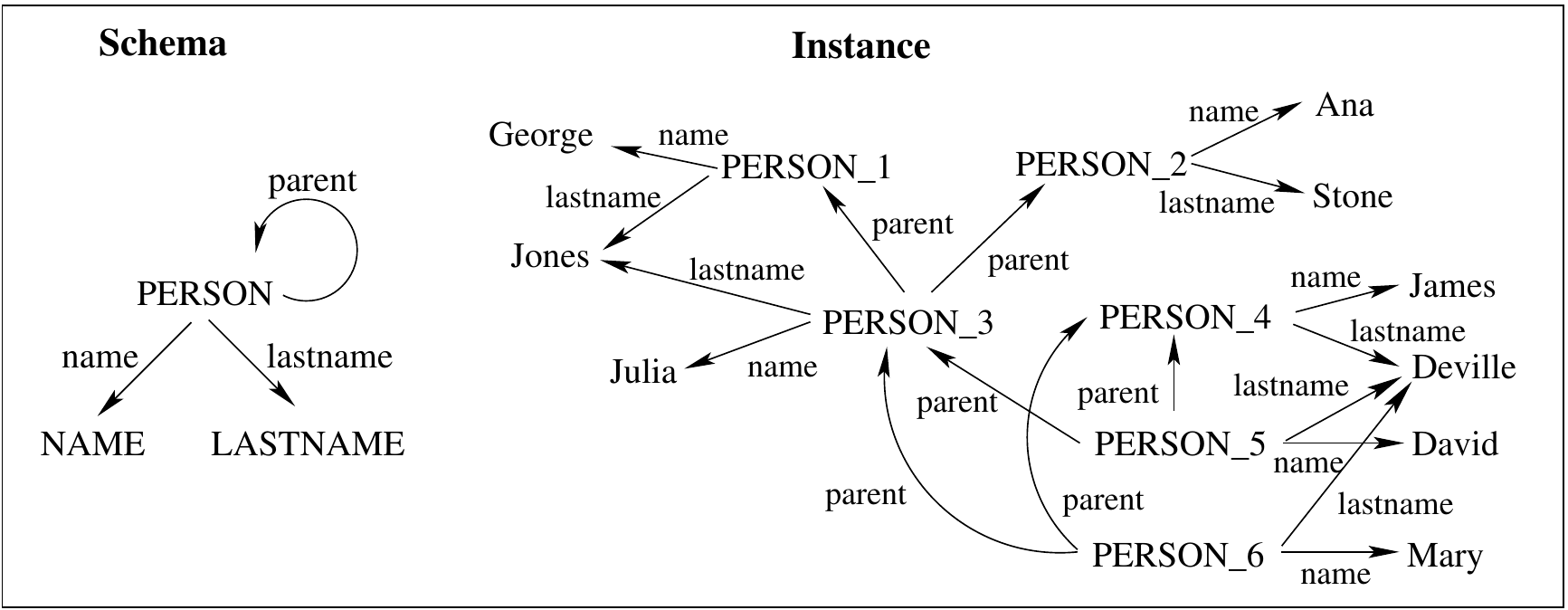}
 \caption{Gram. At the schema level we use generalized names for
 definition of entities and relations. At the instance level, we
 create instance labels (e.g. PERSON\_1) to represent entities, and
 use the edges (defined in the schema) to express relations between
 data and entities.}
 \label{fig:Gram:sample}
\end{figure}

 A schema in Gram is a directed labeled multigraph, where each node is
labeled with a symbol called a \textit{type}, which has associated a domain
of
values.
 In the same way, each edge has assigned a label representing a relation
between types (see example in Figure~\ref{fig:Gram:sample}).
 A feature of Gram is the use of regular expressions for explicit
definition of paths called \textit{walks}.
  An alternating sequence of nodes and edges represent a walk, which
combined with other walks conforms other special objects called
\textit{hyperwalks}.

 For querying the model (particularly path-like queries), an algebraic
language based on regular expressions is proposed.
 For this purpose a hyperwalk algebra is defined, which presents unary
operations (projection, selection, renaming) and binary operations (join,
concatenation, set operations), all closed under the set of hyperwalks.

\subsection{Complex relations: The Hypergraph model}
The notion of hypergraph is 
a generalization of graphs where the notion of edge is extended to {\em hyperedge}, which relates an arbitrary set of nodes~\cite{50216}. 
 Hypergraphs allow the definition of complex objects (using undirected hyperedges), functional dependencies (using directed hyperedges), object-ID and (multiple) structural inheritance.

\begin{figure}[t!]
 \includegraphics[width=12cm]{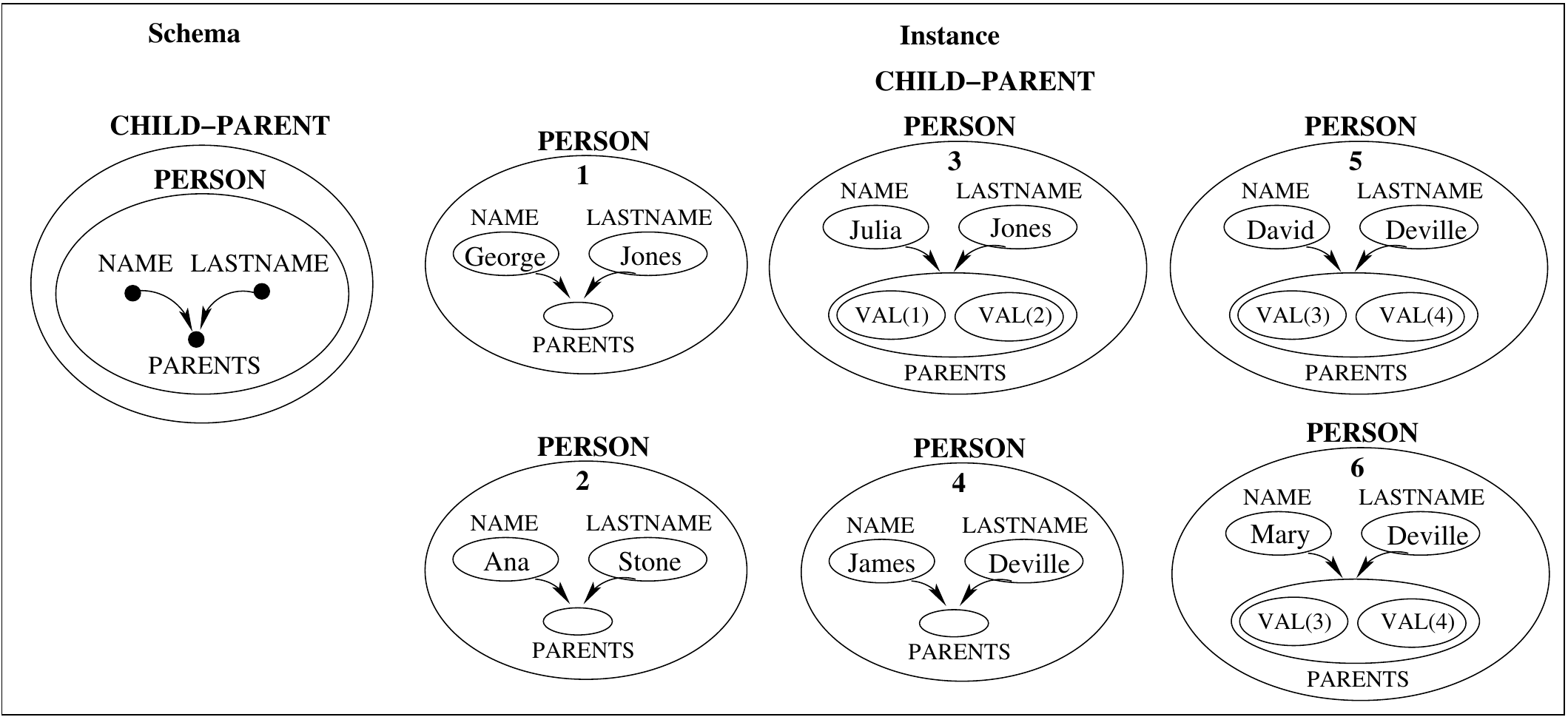}
 \caption{GROOVY. At the schema level (left), we model an object \textit{PERSON} as
 an hypergraph that relates the attributes \textit{NAME},
 \textit{LASTNAME} and \textit{PARENTS}. 
   Note the value functional dependency (VDF)
 \textit{NAME,LASTNAME $\to$ PARENTS} logically represented by the
 directed hyperedge ($\{$NAME,LASTNAME$\}$ $\{$PARENTS$\}$).
 This VFD asserts that NAME and LASTNAME
 uniquely determine the set of PARENTS.}
 \label{fig:groovy:sample}
\end{figure}

  A good representative case is
 GROOVY (Graphically Represented Object-Oriented data model with
 Values~\cite{50187}),  an object-oriented data model which is
 formalized using  hypergraphs. 
 An example of hypergraph schema and instance is presented in Figure~\ref{fig:groovy:sample}.

 The model defines a set of structures for an object data model: value schemas, objects over value schemas, value functional dependencies, object schemas, objects over object schemas and class schemas. 
 The model shows that these structures can be defined in terms of hypergraphs.


Groovy also includes a hypergraph manipulation language (HML) for querying and updating  hypergraphs. 
 It has two operators for querying hypergraphs by identifier or by value, and eight operators for manipulation (insertion and deletion) of hypergraphs and hyperedges.


\subsection{Nested graphs: The Hypernode model}

 A hypernode is a directed graph whose nodes can themselves be graphs (or hypernodes), allowing nesting of graphs. 
Hypernodes can be used to represent \emph{simple} (flat) and 
\emph{complex objects} (hierarchical, composite, and cyclic) as well as mappings and records. 
 A key feature is its inherent ability to \emph{encapsulate information}.

\begin{figure}[h!]
 \centering \includegraphics[width=12cm]{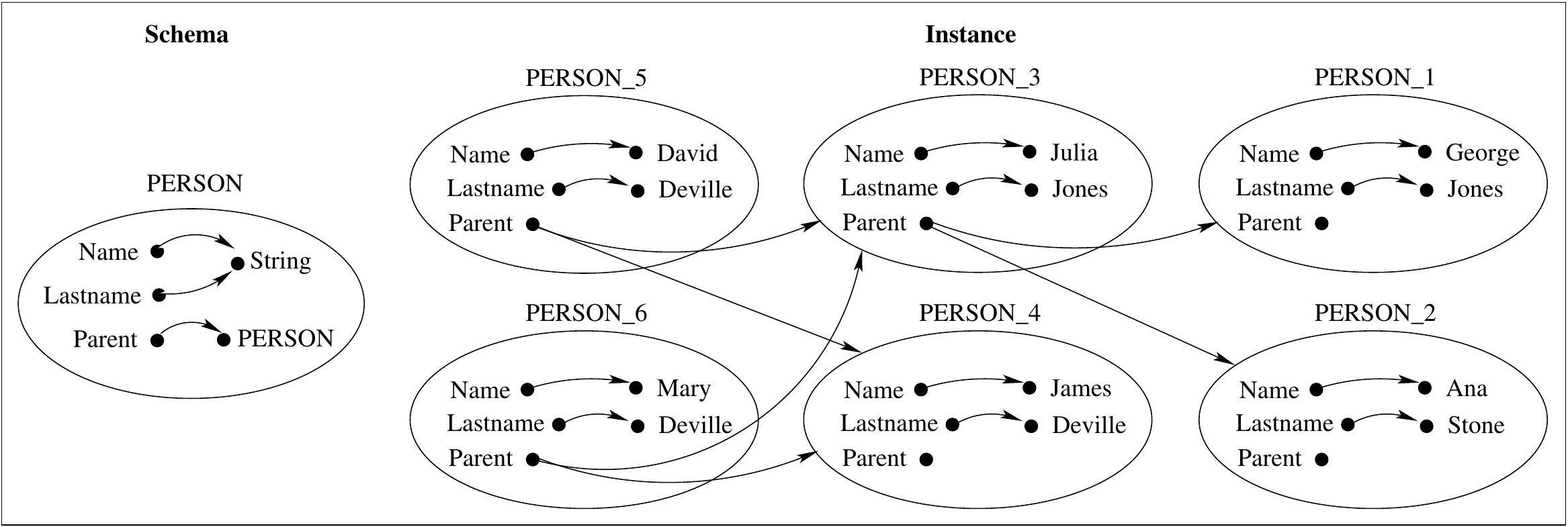}
 \caption{Hypernode Model. The schema (left) defines  a
 \textit{person} as a complex object with the properties
 \textit{name} and \textit{lastname} of type string,
  and \textit{parent} of type person (recursively defined).
    The instance (on the right) shows the relations in the
    genealogy among different instances of person.
    }
 \label{fig:hypernode:sample}
\end{figure}

 The hypernode model which we will use as example 
 was introduced by Levene and Poulovassilis \cite{50167}.
They defined the model and a declarative logic-based language structured as a sequence of instructions (hypernode programs), used for querying and updating hypernodes. 
 A more elaborated version~\cite{50168} includes the notion of schema and type checking, introduced via the idea of types (primitive and complex), that are also represented by nested graphs (See an example in Figure~\ref{fig:hypernode:sample}). 
  It also includes a rule-based query language called \textit{Hyperlog}, which can support both querying and browsing with derivations as well as database updates, and is intractable in the general case.
  A third version of the model~\cite{50152} discusses a set of
  constraints (entity, referential and semantic) over hypernode
  databases.
  In addition it presents another query and update language called HNQL, which use compounded statements to produce HNQL programs.

   Summarizing, the main features of the Hypernode model are:
a nested graph structure which is simple and formal;
the ability to model arbitrary complex objects in a straightforward manner; 
underlying data structure of an object-oriented data model;
enhancement of the usability of a complex objects database system via a graph-based user interface.


\subsection{Trees: The Semistructured model (JSON, OEM, XML)}

 The semistructured model was designed to describe data together with its schema in
one place, also called ``self-describing'' data. Technically they are
trees, the most simple version of a graph, but
 could describe, via references, general graphs.

\begin{figure}[h]
 \centering \includegraphics[width=12cm]{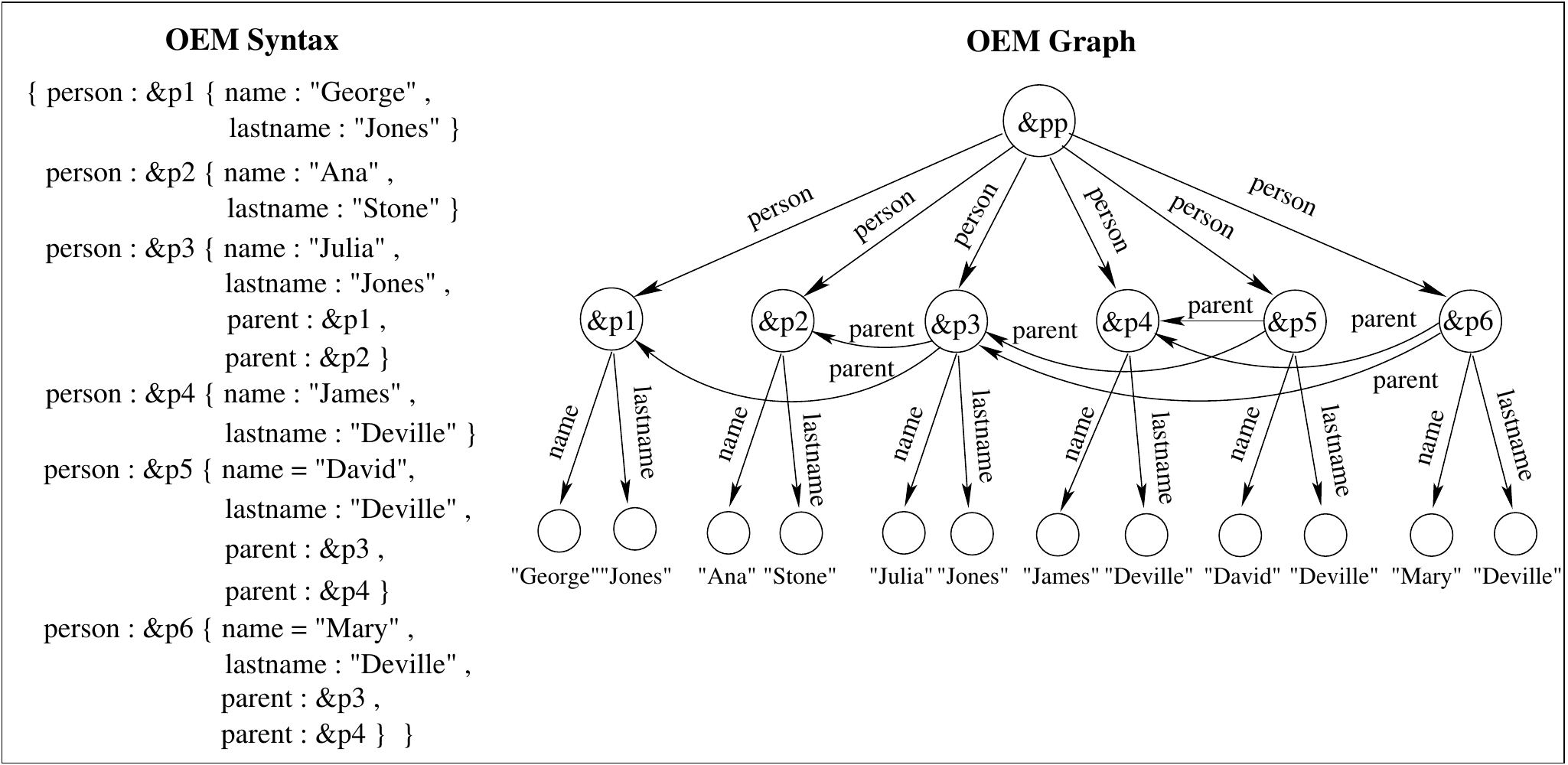}
 \caption{Object Exchange Model (OEM). Schema and instance are mixed.
  The data is modeled  beginning
 in a root node \&pp, with children \textit{person
 nodes}, each of them identified by an Object-ID (e.g. \&p2).
  These nodes have children that contain data
 (\textit{name} and \textit{lastname}) or references to other nodes
 (\textit{parent}). 
   Referencing permits to establish relations between distinct hierarchical
 levels. Note the  tree structure obtained if one 
 forgets the pointers to OIDs, a characteristic of semistructured
 data.}
 \label{fig:OEM:sample}
\end{figure}

The semistructured model was designed to overcome the limitation of 
both, structured data (fixed schema and format, precise rules) and
unstructured data (loose schema, no format, little predictability).
The early motivations were the modeling of documents (whose structure 
can be viewed as trees),  data on the Web and data integration at Web
scale \cite{50103,50122}.

 Among its   advantages are the simple way to integrate new data,
to model incomplete data, and
the flexibility to query it without prior knowledge of schema. The
drawbacks are mainly in the area of optimization, which becomes
much harder as the structure of the data is not necessarily known in advance.

An early proposal in this direction was the data model OEM 
\cite{50409,50155} which proposed an extremely simple and elegant model of objects
with identifiers and ``links'' to other objects , with a simple syntax 
(see Figure \ref{fig:OEM:sample}) which today we can recognize in JSON.

The most popular and elaborated version of the semi-structured model
is the XML model. It comprises a rich and flexible data structure
\cite{XML1}, a suite of highly refined and standardized query and
transformation languages (XPath, XQuery, XSLT)\footnote{
XPath Language \url{www.w3.org/TR/xpath/} \\
XQuery Language \url{www.w3.org/TR/xquery/} \\
XSLT Transformations \url{www.w3.org/TR/xslt20/}
}
and several other
features, that have much to teach graph query language designers.

\subsection{Uniform graphs: The RDF model} 
The Resource Description Framework (RDF)~\cite{10008} is a recommendation of
the W3C designed originally to represent metadata.
One of the main advantages (features) of the RDF model is its ability to
interconnect resources in an extensible way using graph-like structure for
data.

\begin{figure}[h!]
 \includegraphics[width=12cm]{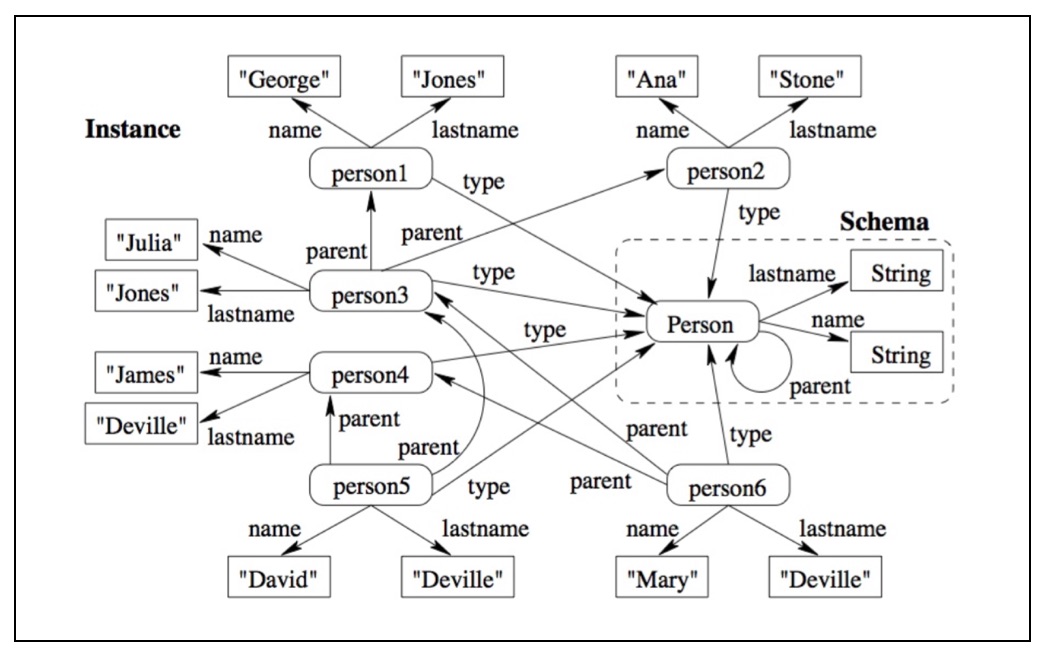}
 \caption{RDF data model. Note that schema and instance are mixed together.
 The edges labeled {\em type} disconnect
 the instance from the schema.  The instance is built
 by the subgraphs obtained by instantiating the nodes
 of the schema, and establishing the corresponding parent edges
 between these subgraphs.
 }
 \label{fig:RDF:sample}
\end{figure}

 One of the main advantages of RDF is its dual nature.
In fact, there are two possible reading of the model.
From a knowledge representation perspective,
 an atomic RDF expression is  triple consisting of a subject (the resource
being described), a predicate (the property) and an object (the property
value).
 Each triple represents a logical statement of a relationship between the subject
and the object, and one could enhance this basic logic by adding rules
and ontologies over it (e.g. RDFS and OWL)
A general RDF expression is a set of such triples called an RDF Graph
(see example in Figure~\ref{fig:RDF:sample}), which can be
intuitively considered as a semantic network.
 From the second perspective, the RDF model is the most general
representation of a graph, where edges are also considered nodes.
In this sense,  formally is not a traditional graph~\cite{10023}. This
allows to self-references, reification (i.e. making statements over
statements), and essentially be self-contained.
 The drawback of all this niceties are the complexity  that 
come with this generalization, particularly for efficient implementation.

SPARQL \cite{10155} is the standard query language for RDF.
It is able to express complex graph patterns by means of a collection of
triple patterns whose solutions can be combined and restricted by using
several operators (i.e. AND, UNION, OPTIONAL, and FILTER).
The latest version of the language, SPARQL 1.1 \cite{90957}, includes
explicit operators to express negation of graph patterns, arbitrary length
path matching (i.e. reachability), aggregate operators (e.g. COUNT),
subqueries, and query federation.


\subsection{Nodes, edges and properties: The Property graph model}
\begin{figure}[t!]
 \centering
 \includegraphics[width=12cm]{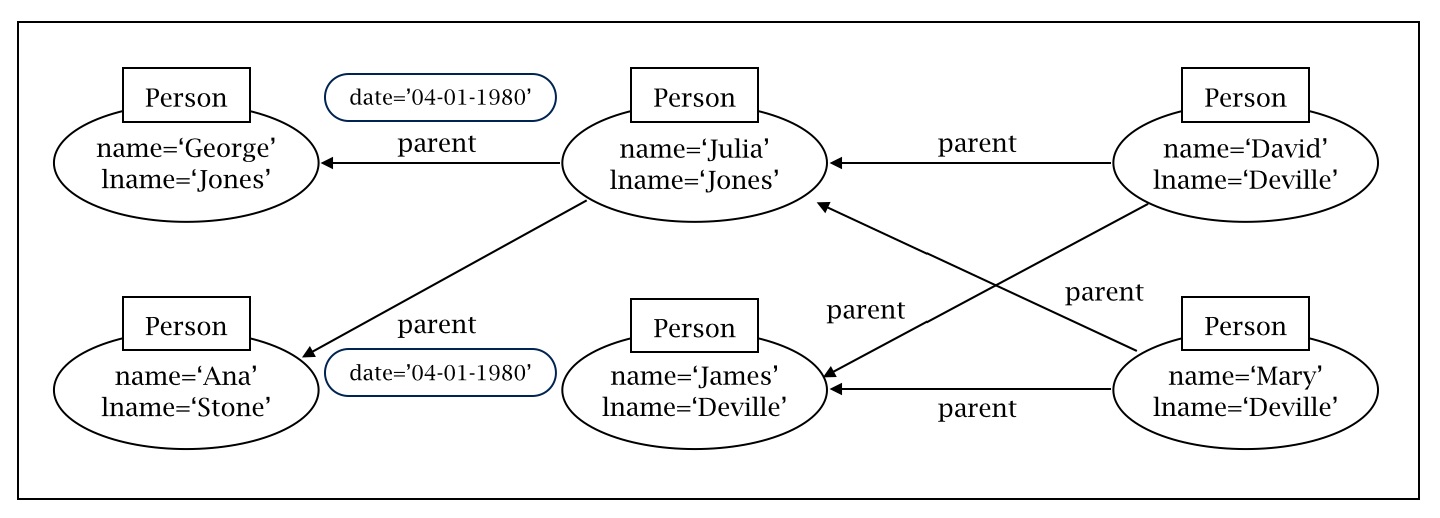}
 \caption{Property graph data model. The main characteristic of this model is the occurrence of properties in nodes and edges. Each property is represented as a pair property-name = ``property-value''.}
 \label{fig:property-graph}
\end{figure}

A \emph{property graph} is a directed, labelled, attributed multigraph.
That is, a graph where the edges are directed, both nodes and edges are labeled and can have any number of properties (or attributes), and there can be multiple edges between any two vertices \cite{91084}.
Properties are key/value pairs that represent metadata for nodes and edges. 
In practice, each vertex of a property graph has an identifier (unique within the graph) and zero or more labels. Node labels could be associated to node typing in order to provide schema-based restrictions.  
Additionally, each (directed) edge has a unique identifier and one or more labels. 
An example of property graph is shown in Figure \ref{fig:property-graph}. 

Property graphs are used extensively in computing as they are more expressive\footnote{Note that the expressiveness of a model is defined by ease of use, not by the limits of what can be modeled.} than the simplified mathematical objects studied in theory.
In fact, the property graph model can express other types of graph models by simply abandoning or adding particular bits and pieces \cite{91084}.

There is no standard query language for property graphs although some proposals are available.
 Blueprints \cite{91154} was one of the first libraries created for the property graph data model. Blueprints is analogous to the JDBC, but for graph databases. 
 Gremlin \cite{91146} is a functional graph query language which allows to express complex graph traversals and mutation operations over property graphs. 
 Neo4j \cite{90278} provides Cypher \cite{91040}, a declarative query language for property graphs.
 The syntax of Cypher, very similar to SQL via expressions match-where-return, allows to easily express graph patterns and path queries. 



\section{Querying Graph Databases}

Data manipulation and querying in graph data management is expressed by graph operations or graph transformations whose main primitives are based on graph features like neighborhoods, graph patterns and paths. 
 Another approach is to express all queries using a few powerful graph manipulation primitives enclosed by a graph query language. 

 In this section we give a brief overview of the research on querying graph databases.
 First, we present a broad classification of queries studied in the context of graph databases, including a description of their characteristics (e.g. complexity and expressiveness). After that, we present a review of graph query languages, including short descriptions of some proposals we consider representative of the area.  
 

\subsection{Classification of graph queries} 

 In this section we present a broad classification of queries that have been largely studied in graph theory and can be considered essential for graph databases. 
 We grouped them in adjacency, pattern matching, reachability and analytical queries.
  
To fix notations, let us represent a graph database as a single labeled directed multigraph.
Specifically, a tuple $G=(N,E,L,\delta,\lambda_N,\lambda_E)$ where 
$N$ is a finite set of nodes, 
$E$ is a finite set of edges, 
$L$ is a finite set of labels,
$\delta : E \to N^2$ is the edge function that associates edges with pairs of nodes, 
$\lambda_N : N \to L $ is the node labeling function, and 
$\lambda_E : E \to L $ is the edge labeling function.
 An edge $e = (n,n') \in E$ will be represented as a triple $(v,w,v')$ 
 where $v = \lambda_N(n)$, $w = \lambda_E(e)$ and $v' = \lambda_N(n')$.
 Nodes and edges will usually be referenced by using their labels.   
 Additionally, a path $\rho$ in $G$ is a sequence of edges 
$(v_0,w_0,v_1)$, $(v_1,w_1,v_2)$, \dots, $(v_{m-1},w_{m-1},v_m)$ 
where $v_0$ and $v_m$ are the source and target nodes of the path respectively.
 The label of $\rho$ is the sequence of labels $w_0, w_1, \dots, w_{m-1}$. 
  

\subsubsection{Adjacency queries}
The primary notion in this type of queries is node/edge adjacency. 
Two nodes are adjacent (or neighbors) when there is an edge between them. 
Similarly, two edges are adjacent when they share a common node.
 Examples of adjacency queries are: ``return the neighbors of a node $v$'' or ``check whether nodes $v$ and $v'$ are adjacent'',
 In spite of their simplicity, to compute efficiently adjacency queries could be a challenge for big sparse graphs \cite{90452}. 

  
 The basic notion of adjacency can be extended to define more complex ``neighborhood queries''.
 For instance, the k-neighborhood \cite{70083} of a root node $v$ is the set of all nodes that are reachable from $v$ via a path of $k$ edges (i.e. the length of the path is no more than $k$).   
 Similarly, the k-hops \cite{90386} of $v$ returns all the nodes that are at a distance of $k$ edges from $v$.
 Note that a k-neighborhood query can be expressed as a composition of k-hops queries 1-hops $\cup \dots \cup$ k-hops, but removing duplicates. 

 Several applications can benefit from reachability queries, in particular those where the notion of influence is an important concern.  
 For instance, in information retrieval adjacency queries are used for web ranking using hubs and authorities \cite{90410}.  
 In recommendation systems are used to obtain users with similar interests \cite{90386}.
 In social networks can be used to validate the well-known six-degrees-of-separation theory.



\subsubsection{Pattern matching queries}
The basic notion of graph pattern matching consists in to find the set of subgraphs of a database graph that ``match'' a given graph pattern.
 A basic graph pattern is usually defined as a small graph where some nodes and edges can be labeled with variables. The purpose of the variables is to indicate unknown data and more importantly, to define the output of the query (i.e. variables will be ``filled'' with solution values).  
 For instance, the expression $(John,friend,?y)$, $(John,friend,?z)$, $(?y,friend,?z)$ 
 represents a graph pattern where $?x$ and $?y$ are variables.
 The result or interpretation of this graph pattern could be ``the pairs of friends of John which are also friends''. 
  
 Graph pattern matching is typically defined in terms of sub-graph isomorphism, i.e. to find all subgraphs of a database $G$ that are isomorphic to a graph pattern $P$. 
 Hence, pattern matching deals with two problems: the graph isomorphism problem that has a unknown computational complexity, and the sub-graph isomorphism problem which is an NP-complete problem~\cite{91195}.
 
Graph matching is easily identifiable in many application domains.
For instance, graph patterns are fundamental within the pattern recognition field \cite{91268}.
In social network analysis is used to identify communities and social positions \cite{91226}.
In protein interaction networks, researchers are interested in patterns that determine proteins with similar functions \cite{91269}.
  
There are a number of variations on the basic notion of pattern matching:
\begin{itemize}
\item \emph{Graph patterns with structural extension or restrictions.} 
A basic graph pattern has been defined as a simple structure containing nodes, edges and variables, however this notion can be extended or restricted depending on the graph data model.
 For instance, if the database is a property graph then a graph pattern should support conditions over such properties.        
\item \emph{Complex graph patterns.} 
In some cases, a collection of basic graph patterns can be combined via specific operators (e.g. union, optional and difference) to conform complex graph patterns. 
 The semantics of these graph patterns can be defined in terms of an algebra of graph patterns.
\item \emph{Semantic matching.} 
It consists in to match graphs based on specific interpretations (i.e. semantics) given to nodes and edges. Such interpretations can be defined via semantic rules (e.g. an ontology).
\item \emph{Inexact matching.} 
In this case the graph pattern matching algorithm returns a ranked list of the most similar matches (instead of the original exact matching). 
These algorithms employ a cost function to measure the similarity of the graphs and error correction techniques to deal with noise data. 
\item \emph{Approximate matching.}
This variation concerns the use of algorithms that find approximate solutions to the pattern matching problem, i.e. they offer polynomial time complexity but are not guaranteed to find a solution.
 In case of exact matching the algorithm will return some solutions, but not all matches. For inexact matching, a close solution will be returned, but not the closest. 
\end{itemize}

Very related to graph pattern matching is the area of \emph{graph mining} \cite{90413}. This area includes the problems of frequent pattern mining, clustering and classification.  
 For instance, the goal of frequent pattern mining is the discovery of common  patterns, i.e.  to find subgraphs that occurs frequently in the entire database graph.
 The problem of computing frequent subgraphs is particularly challenging and computationally intensive, as it needs to compute graph and subgraph isomorphisms.    
 The discovery of patterns can be useful for many application domains, including finding strongly connected groups in social networks and finding frequent molecular structures in biological databases.

\subsubsection{Reachability queries (connectivity)}
 One of the most characteristic problems in graph databases is to compute reachability of information. In general terms, the problem of reachability tests whether two given nodes are connected by a path.
  Reachability queries have been intensively studied in traditional database models, in particular for querying relational and semi-structure databases.
Yannakakis \cite{50065} surveyed a set of path problems relevant to the database area including computing transitive closures, recursive queries and the complexity of path searching.

 In the context of graph databases, reachability queries are usually modeled as path or traversal problems characterized by allowing restrictions over nodes and edges.    
 Mendelzon and Wood \cite{50158} introduced the notion of Regular Path Query (RPQ) as a way of expressing reachability queries.
 The basic structure of a regular path query is an expression $(?x,\tau,?y)$ where $?x$ and $?y$ are variables, and $\tau$ is a regular expression.
 The goal of this RPQ is to find all pairs of nodes $(?x,?y)$ connected by a path such that the concatenation of the labels along the path satisfies $\tau$.
 Note that variables $?x$ and $?y$ can be replaced by node labels (i.e. data values) in order to define specific source and target nodes respectively.
 For instance, the path query $(John,friend^+,?z)$ returns the people $?z$ that can be reached from ``John'' by following ``friend'' edges.   

 The complex nature of path problems is such that their computations often requires a search over a sizable data space.    
 The complexity of regular path queries was initially studied in \cite{91224} in terms of computing simple paths (i.e. paths with no repeated nodes). Specifically, the problem of finding all pairs of nodes connected by a simple path satisfying a given regular expression was shown to be NP-complete in the size of the graph. 
 Due to the high computational complexity of RPQs under simple path semantics, researchers proposed a semantics based on arbitrary paths. 
 This semantics leads to tractable combined complexity for RPQs and tractable data complexity for a family of expressive languages. 
 See \cite{90852} for a complete review about these issues.
 
 Reachability queries are present in multiple application domains.
 For instance, path queries are very useful: 
in semi-structured data, for querying XML documents using XPath \cite{91254};  
in social networks, to discover people with common interests \cite{91226}; 
and
in biological networks, to find specific biochemical pathways between distance nodes \cite{91269}.

Reachability queries are the basis for other real life graph queries.
Maybe the most important is the \emph{shortest-path distance} (also called the geodesic distance).
For instance, in a road network it is fundamental to calculate the minimum distance  between two locations \cite{90758}.

\subsubsection{Analytical queries}
The queries of this type do not consult the graph structure; instead they are oriented to measure quantitatively and usually in aggregate form topological features of the database graph.
 Analytical queries can be supported via special operators that allow to summarize the query results, or by ad-hoc functions hiding complex algorithms.

Summarization queries can be expressed in a query language by using the so-called aggregate operators (e.g., average, count, maximum, etc.).
These operators can be used to calculate 
the order of the graph (i.e., the number of vertices), 
the degree of a node (i.e., the number of neighbors of the node), 
the minimum / maximum / average degree in the graph, 
the length of a path (i.e., the number of edges in the path),  
the distance between nodes (i.e., the length of a shortest path between the nodes), 
among others ``simple'' analytical queries.

 Complex analytical queries are related to compute important algorithms for graph analysis and mining  (see \cite{90413} for a extensive review).
 Examples of such graph algorithms are:
\begin{itemize}
\item  \emph{Characteristic path length}. It is the average shortest path length in a network. It measures the average degree of separation between the nodes. 
\item 
\emph{Connected components}. It is an algorithm for extracting groups of vertices that can reach each other via graph edges.
\item 
 \emph{Community detection}. This algorithm deals with the discovery of groups whose constituent nodes form more relationships within the group than with nodes outside the group. 
\item
\emph{Clustering coefficient}. The clustering coefficient of a node is the probability that the neighbors of the node are also connected to each other. The average clustering coefficient of the whole graph is the average of the clustering coefficients of all individual nodes.
\item 
\emph{PageRank} 
This algorithm, created in the context of web searching, models the behavior of an idealized random Web surfer.
The PageRank score of a webpage represents the probability that the random Web surfer chooses to view the webpage.
This algorithm can be an effective method to measure the relative importance of nodes in a data graph. 
\end{itemize}

 Complex analytical queries are the speciality of graph processing frameworks due to their facilities for implementing and running complex algorithms over large graphs.
 More details about these queries can be found in articles comparing graph processing frameworks (e.g. \cite{91093} and \cite{91090}).

\subsection{A short review of graph query languages} 

 In the literature of graph data management there is substantial work on graph query languages (GQLs).
 A review of GQLs proposed during the first wave of graph databases was presented in \cite{90004}. 
 Based on this, Wood \cite{90524} studied several GQLs focusing on their expressive power and computational complexity.
 A review and comparison of practical query languages provided by graph databases (available at the time) was presented in \cite{90525}.  
 Recently, Barcelo \cite{90852} studied the problem of querying graph databases, in particular the expressiveness and complexity of several navigational query languages.
 
 To the best of our knowledge, the first graph query language was proposed in 1987 by Cruz, Mendelzon and Wood (i.e. the language G). After that, several theoretical GQLs have been proposed, in some cases accompanying the definition of a graph data model. 
 It has been just in the last ten years that practical GQLs are available thanks to the release of graph database systems. Example of this is Cypher, the query language provided by the Neo4j graph database.    
   
 Although a GQL is normally related to a graph database model, this relation is no exclusive.
 For instance, several object-oriented data models defined graph-based languages to manipulate the objects in the database (e.g. GraphDB \cite{50339} and G-Log\cite{50081}), or to represent database transformations (e.g. GOOD \cite{50073} and GUL \cite{50298}).
 A similar situation occurred for semistructured data models when graph-oriented operations are used to navigate the tree-based data (e.g. Lorel \cite{50154} and UnQL \cite{90693}).
 Additionally, several graph-based query languages have been designed for specific applications domains, in particular those related to complex networks, for instance social networks (e.g. SoQL \cite{90698}), biological networks (e.g. DNAQL \cite{91180}), bibliographical networks (e.g. BiQL \cite{90446}), the Web (e.g. WebSQL \cite{90446}) and the Semantic Web (e.g. SPARQL \cite{90699}).    
 
\begin{figure}[t!]
 \centering
 \includegraphics[width=12cm]{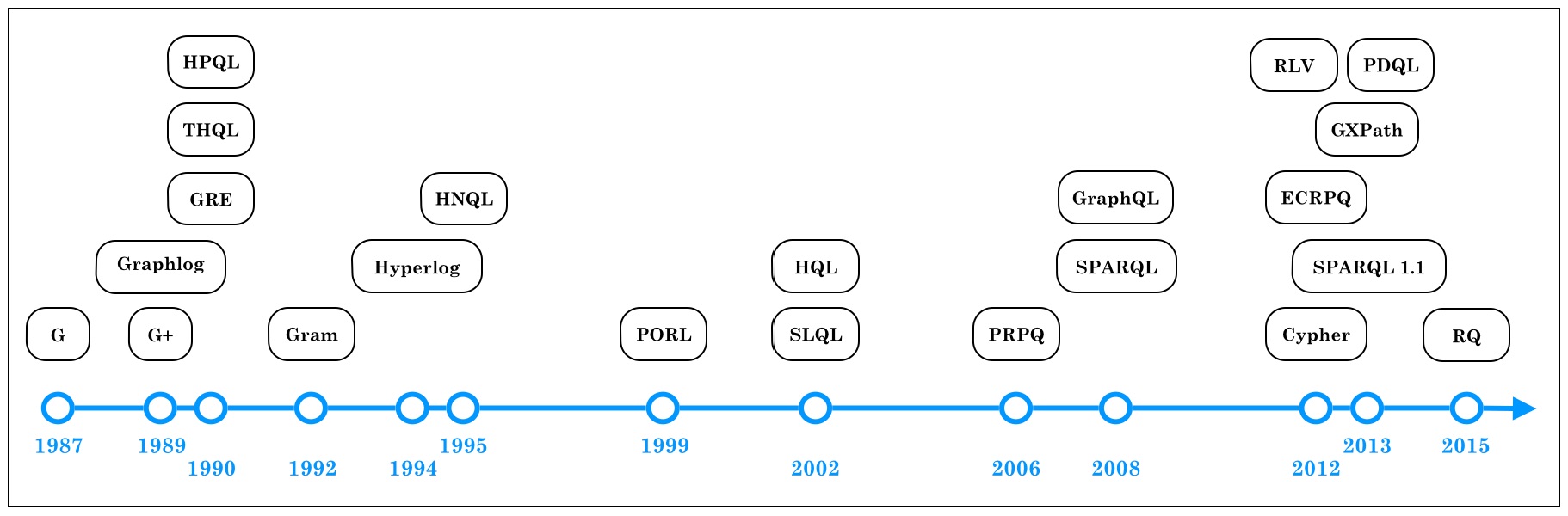}
 \caption{Evolution of graph query languages: G \cite{50158}, G+ \cite{50285}, Graphlog \cite{50079}, HPQL \cite{50167}, THQL \cite{50188}, GRE \cite{91251}, Gram \cite{50019}, Hyperlog \cite{50168}, HNQL \cite{50529}, PORL \cite{50250}, SLQL \cite{50064}, HQL \cite{50659}, PRPQ \cite{90405}, GraphQL \cite{90169}, SPARQL \cite{10155}, RLV \cite{91244}, Cypher \cite{91040}, ECRPQ \cite{91257}, PDQL \cite{90825}, GXPath \cite{91243}, SPARQL 1.1 \cite{90957} and RQ \cite{91245}.}
 \label{fig:gqls}
\end{figure}

  For the sake of space we will not present a complete review of graph query languages. Instead we describe some of the languages we consider relevant and useful to show the developments in the area.
 Moreover, we restrict our review to ``pure'' GQLs, that is those languages specifically designed to work with graph data models.
 Figure \ref{fig:gqls} presents this subset of languages in chronological order.  

\begin{figure}[t!]
 \centering
 \includegraphics[width=10cm]{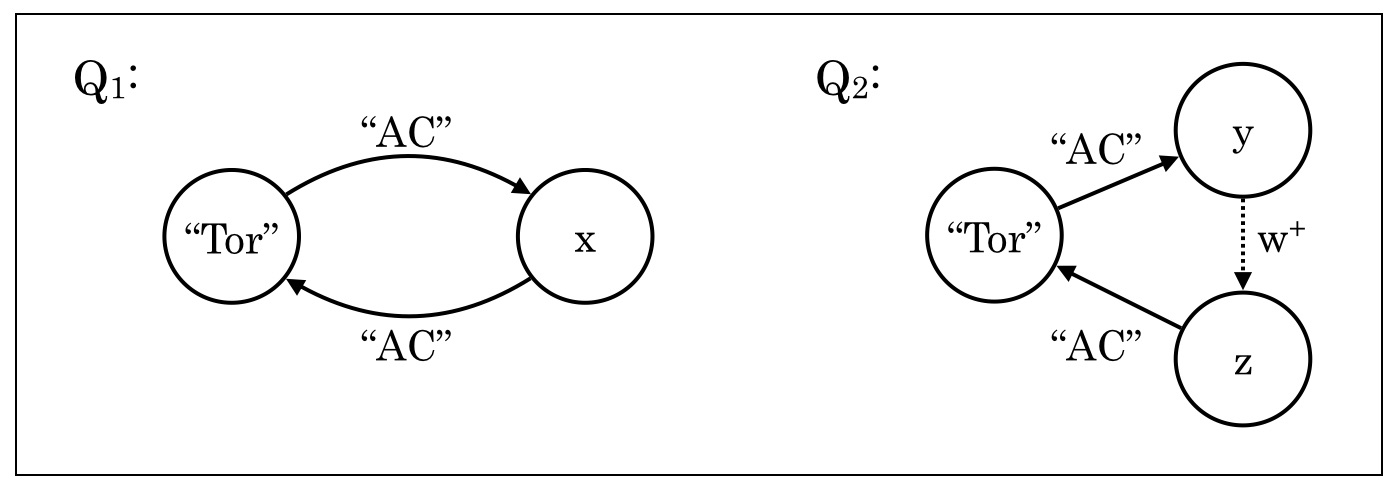}
 \caption{Example of a graphical graph query expressed in language G \cite{50158}.}
 \label{fig:g-query}
\end{figure}

 As we mentioned before, Cruz et al.~\cite{50158} proposed the query language G.
 This language introduced the notion of graphical query as a set of query graphs.
 A query graph (pattern) is a labeled directed multigraph in which the node labels may be either variables or constants, and the edge labels can be regular expressions combining variables and constants.
 The result of a graphical query $Q$ with respect to a graph database $G$ is the union of all query graphs of $Q$ which match subgraphs of $G$. 
 For instance, Figure \ref{fig:g-query} presents a example of graphical query containing two query graphs, $Q_1$ and $Q_2$. This query finds the first and last cities visited in all round trips from Toronto (``Tor''), in which the first and last flights are with Air Canada (``AC'') and all other flights (if any) are with the same airline.
  Note that the last condition is expressed by the edge labeled with regular expression $w^+$.    
 Thanks to the inclusion of regular expressions, G is able to express recursive queries more general than transitive closure.
 However, the evaluation of queries in G is of high computational complexity due to its semantics based on simple paths.   
  
 G evolved into a more powerful language called G+~\cite{50285}. The notion of graphical query proposed by G is extended in G+ to define a summary graph that represent how to restructure the answer obtained by the query graphs. Additionally, G+ allows to express aggregate functions over paths and sets of paths (i.e. it allows to compute the size of the shortest path).

 GraphLog~\cite{50209} is a query language that extends G+ by adding negation and unifying the concept of a query
graph. A query is now a single graph pattern containing one distinguished edge which corresponds to the restructured edge of the summary graph in G+. 
 The effect of a GraphLog query is to find all instances of the pattern that occur in the database graph and for each one of them define a virtual link represented by the distinguished edge.
 It was shown \cite{50079} that the expressive power of GraphLog is equivalent to three well-known query classes: stratified linear Datalog programs, queries computable in non-deterministic logarithmic space, and queries expressible with a transitive closure operator plus first-order logic. Based on this, the GraphLog's authors argued that the language is able to express ``real life'' recursive queries. 

Gram~\cite{50019} is a query language based on walks\footnote{In graph theory, a walk is an alternating sequence of nodes and connecting edges, which begins and ends with a vertex, and where any node and any edge can be visited any number of times.} and hyperwalks.
Assuming that $T$ is the union of node and edge types in the database graph, a walk expression is a regular expression over $T$ without alternation (union), whose language contains only alternating sequences of node and edge types.
A hyperwalk is a set of walk expressions connected by at least one node type. 
Assuming a database graph containing travel agency data, the expression
\texttt{JOURNEY first (STOP next)* + STOP in CITY}
is a hyperwalk containing two walk expressions connected by the node type STOP. 
Hence, the above hyperwalk describes the walks going from a node (of type) JOURNEY to one of its nodes (of type) STOP in a CITY.

The set of walks in the database satisfying a hyperwalk expression $r$ is called the instance of $r$ and is denoted by $I(r)$.  
Based on these notions, Gram defines a hyperwalk algebra with operations closed under the set of hyperwalks (e.g. projection, selection, join and set operations).
For example, the following algebra expression computes all journeys which traverse Munich.
\begin{center}
$\pi_{JOURNEY}(\sigma_{Munich(CITY)}$I(JOURNEY~first(STOP~next)* STOP~in~CITY))) 
\end{center}

Although less popular, there are also languages for manipulating and querying hypergraphs and hypernodes (nested graphs). GROOVY~\cite{50187} introduced a Hypergraph Manipulation Language (HML) for querying and updating labeled hypergraphs which defines basic operators for manipulation (addition and deletion) and querying of hypergraphs and hyperedges. On the other side, Levene and Poulovassilis \cite{50167} defined a logic-based query and update language for hypernode where a query is expressed as a hypernode program consisting of a set of hypernode rules. 

GraphQL \cite{90169} is a graph query language for property graphs which is based on the use of formal grammars for composing and manipulating graph structures.
A graph grammar is a finite set of graph motifs where a graph motif can be either a simple graph or composed of other graph motifs by means of concatenation, disjunction and repetition.
For instance, consider the following graph grammar containing three graph motifs:
\begin{quote}
\textbf{graph} $G_1$ \{ \textbf{node} $v_1$, $v_2$; \textbf{edge} $e_1$($v_1$,$v_2$); \}
\\
\textbf{graph} $G_2$ \{ \textbf{node} $v_2$, $v_3$; \textbf{edge} $e_2$($v_2$,$v_3$); \}
\\
\textbf{graph} $G_3$ \{ \textbf{graph} $G_1$ as X; \textbf{graph} $G_2$ as Y; \textbf{edge} $e_3$(X.$v_2$, Y.$v_2$) \}.
\end{quote}
The graph motifs $G_1$ and $G_2$ are simple, whereas $G_3$ is a complex graph motif which concatenates the graph motifs $G_1$ and $G_2$ via the edge $e_3$ and the common node $v_2$.
The language of a graph grammar is the set of all the graphs derivable from graph motifs of that grammar.
 The query language is based on graph patterns consisting of a graph motif plus a predicate on attributes of the motif. A predicate is a combination of boolean or arithmetic comparison expressions.
 For instance, the expression 
\begin{quote}
\textbf{graph} $P$ \{  \textbf{node} $v_1$, $v_2$; \textbf{edge} $e_1$($v_1$,$v_2$) \}
\\ 
\textbf{where} $v_1$.name=``A'' and $v_2$.year $>$ 2000;    
\end{quote}
describes a graph pattern where two nodes $v_1,v_2$ must be connected by an edge $e_1$, and the nodes must satisfy the conditions following the \textbf{where} clause. 

Note that most of the languages described above are more theoretical than practical. 
Cypher \cite{91040} is a declarative language for querying property graphs implemented by the Neo4j graph database.
The most basic query in Cypher consists of a expression containing clauses START, MATCH and RETURN. 
For example, assuming a friendship graph, the following query returns the name of the friends of the persons named ``John'':
\begin{verbatim}
START x=node:person(name="John")
MATCH (x)-[:friend]->(y)
RETURN y.name
\end{verbatim}
The START clause specifies one or more starting points (nodes or edges) in the database graph.
The MATCH clause contains the graph pattern of the query.
The RETURN clause specifies which nodes, edges and properties in the matched data will be returned by the query.
Cypher is able to express some types of reachability queries via path expressions. 
For instance, the expression \texttt{p = (a)-[:knows*]->(b)} computes the paths from node \texttt{(a)} to node \texttt{(b)}, following only \texttt{knows} outgoing edges, and maintains the solution in the path variable \texttt{p}. 
Additionally, there exist build-in functions to calculate specific operations on nodes, edges, attributes and paths. 
For instance, complementing the above path expression, the function \texttt{shortestPath(p)} returns the shortest path between nodes \texttt{(a)} and \texttt{(b)}.  


SPARQL \cite{10155} is the standard query language for the RDF data model.
A typical query in SPARQL follows the traditional SELECT-FROM-WHERE structure where the FROM clause indicates the data sources, the WHERE clause contains a graph pattern, and the SELECT clause defines the output of the query (e.g. resulting variables).    
The simplest graph pattern, called a triple pattern, is an expression of the form 
$subject-predicate-object$ where identifiers (i.e. URIs), values (RDF Literals) or variables (e.g. ?X) can be used to represent a node-edge-node pattern.
A complex graph pattern is a collection of triple patterns whose solutions can be combined and restricted by using operators like AND, UNION, OPTIONAL and FILTER.
For instance, the following query returns the names of persons described in the given data source (i.e. an RDF graph): 
\begin{verbatim}
SELECT ?N 
FROM <http://example.org/data.rdf> 
WHERE { ?X rdf:type voc:Person . ?X voc:name ?N } 
\end{verbatim}
The latest version of the language, SPARQL 1.1 \cite{90957}, includes novel features like negation of graph patterns, arbitrary length path matching (i.e. reachability), aggregate operators (e.g. COUNT), subqueries, and query federation.

\section{Graph Data Management Systems}
\label{systems}

The systems for graph data management can be classified in two main categories, graph databases and graph processing frameworks. 
Although the problems addressed for both groups are similar, they provide two different approaches for storing and querying graph data, with their own advantages and disadvantages.

 Graph databases aim at persistent management of graph data, allowing to transactionally store and access graph data on a persistent medium. In this sense, these provide efficient single-node solutions with limited scalability.  
 On the other hand, graph processing frameworks aim to provide batch processing and analysis of large graphs often in a distributed environment with multiple machines. These solutions usually process the graph in memory, but different parts of the graph are managed by distinct, distributed nodes. 
 
 Closely related to graph databases are the systems for managing RDF data.
These systems, called RDF Triple Stores or RDF databases, are specifically designed to store collections of RDF triples, to support the standard SPARQL query language, and possibly to allow some kind of inference via semantic rules.
 Although Triple Stores are based on the RDF graph data model, they are specialized databases with their own characteristics. 
 Therefore, we will study them separately.
  
 Next we present a review of current systems in the above categories, including a short description of each of them.

\subsection{Graph database systems}

 A graph database system (GDBS) -- or just graph database  -- is a system specifically designed for managing graph-like data following the basic principles of database systems, i.e. persistent data storage, physical/logical data independence, data integrity and consistency.  
 The research on graph databases have a long history, at least since the 1980’s.
 Although the first of these were primarily theoretical proposals (with emphasis on graph database models), it is only recently that several technological developments (e.g. powerful hardware to store and process graphs) have made it possible to have practical systems.


 The current ``market'' of graph databases includes more than 15 systems with different levels of development. 
 All these systems provide most of the major components in database management systems, including:
storage engine (with innate support for graph structures),
database languages (for data definition, manipulation and querying),
indexes and query optimizer,
transactions and concurrency controllers, 
and
external interfaces (user interface or API) for system management.
 
 Considering their internal implementation, we classify graph databases in two types: native and non-native graph databases. 
 Native graph databases implement ad-hoc data structures and indexes for storing and querying graphs. 
 Non-native graph databases make use of other database systems to store graph data and implement query interfaces to execute graph queries over the back-end system.    

 In the group of native graph databases we found:
AllegroGraph \cite{90283}, 
Bitsy \cite{91076},
Cayley \cite{91018},  
GraphBase \cite{90432},
Graphd \cite{91098},
HyperGraphDB \cite{90402}, 
IBM System G \cite{91103},
imGraph \cite{91074}, 
InfiniteGraph \cite{90434}, 
InfoGrid \cite{90281},
Neo4j \cite{90278}, 
Sparksee/DEX \cite{91022}, 
Trinity \cite{90749})
and
TurboGraph \cite{91041}. 
 Among the systems considered non-native graph databases we can mention:
 Titan \cite{91023}, which supports Apache Cassandra, Apache HBase and Oracle BerkeleyDB as storage backends;
 FlockDB \cite{90488} is a distributed graph-oriented database which uses MySQL as the storage engine;
OrientDB \cite{91111} and ArangoDB \cite{90718}, which are document-store databases adapted to managing graphs; 
OQGRAPH \cite{91112} which is a graph computation engine for MySQL, MariaDB y Drizzle;    
 VelocityGraph \cite{91119} an object database supporting graphs; and
 Horton \cite{91100}, which is based on the cloud programming infrastructure Orleans.


There are several papers comparing the 
features \cite{90400,90525,90783,91057} and 
performance \cite{90264,90672,91129} 
of graph databases.
 Next, we briefly describe the systems we consider more relevant. 

AllegroGraph\cite{90283} is one of the precursors in the current generation of graph databases. 
Although it was born as a graph database, its current development is oriented to meet the Semantic Web standards (i.e., RDF/S, SPARQL and OWL).
 Additionally, AllegroGraph provides special features for GeoTemporal Reasoning and Social Network Analysis.

Sparksee (formely DEX)\cite{90080} is a native graph database for persistent storage of property graphs.
 Its implementation is based on bitmaps and other secondary structures, and provides libraries (APIs) in several languages for implementing graph queries.
 Sparksee is been used in social, bibliographical and biological networks analysis, media analysis, fraud detection and business intelligence applications of indoor positioning systems

HyperGraphDB~\cite{90402} is a systems that implements the hypergraph data model (i.e. edges are extended to connect more than two nodes). 
 This model allows a natural representation of higher-order relations, and is particularly useful for modeling data of areas like knowledge representation, artificial intelligence and bio-informatics. 
 Hypergraph stores the graph information in the form of key-value pairs	which are stored on BerkeleyDB. 


InfiniteGraph~\cite{90434} is a database oriented to support large-scale graphs in a distributed environment.
 It aims the efficient traversal of relations across massive and distributed data stores.
 Its focus of attention is to extend business, social and government intelligence with graph analysis.

Neo4j~\cite{90278} is based on a network oriented model where relations are first class objects.
It is fully written in java and implements an object-oriented API, a native disk-based storage manager for graphs, and a framework for graph traversals.

Trinity \cite{90749}) implements a general purpose graph engine over a distributed memory cloud.
Trinity implements a globally addressable distributed memory storage, and provides a random access abstraction for large graph computation.
Hence, it supports both online graph query processing and offline graph analytics.
Its query languages, called TSL, allows users to declare data schema and communication protocols.

\subsection{Graph processing frameworks}

In addition to graph databases, a number of graph processing frameworks have been proposed to address the needs of processing complex and large-scale graph datasets.
These frameworks are characterized by in-memory batch processing and the use of distributed and parallel processing strategies.
 Note that, distributed systems with more computing and memory resources are able to process large-scale graphs, but they can be less efficient than single-node platforms when specific graph queries are executed.



 On the one hand, generic data processing systems such as Hadoop \cite{91147}, YARN \cite{91148}, Stratosphere \cite{91149} and Pegasus \cite{91048} have been adapted for graph processing due to their facilities for batch data processing.
 Most of these systems are based on the MapReduce programming model and implemented on top of the Hadoop platform, the open source version of MapReduce.  
 By exploiting data-parallelism, these systems are highly scalable and support a range of fault-tolerance strategies.
 Though these systems improve the performance of iterative queries, users still need to ``think'' their analytical graph queries as MapReduce jobs.
In fact, naively expressing graph computation and graph algorithms in these data-parallel abstractions can be challenging \cite{91132}.
 Additionally, these systems cannot take advantage of the characteristics of graph-structure data and often result in complex job chains and excessive data movement when implementing iterative graph algorithms \cite{91090}. 

 On the other hand, graph-specific platforms such as 
Pregel \cite{90266}, 
Apache Giraph \cite{91061}, 
GraphLab \cite{90830,91044}, 
Apache Hama, 
Catch de Wind \cite{91087}, 
GPS \cite{91081},
Mizan \cite{91086},
PowerGraph \cite{91045},
GraphX \cite{91058},
TurboGraph \cite{91041}
and 
GraphChi \cite{91047}
provide different programming interfaces for expressing graph analytic algorithms.
 These platforms, also called \emph{offline graph analytic systems}, perform an iterative, batch processing over the entire graph dataset until the computation satisfies a fixed-point or stopping criterion.
 Therefore, these systems are particularly designed for computing graph algorithms which require iterative, batch processing, e.g., PageRank, recursive relational queries, clustering, social network analysis, machine learning and data mining algorithms \cite{91125}.
  Next we briefly describe some of these systems. 



 Pregel \cite{90266} is a system that provides a native API specifically designed by Google for writing algorithms that process graph data.
 Pregel is a vertex-centric programming abstraction that adapts the Bulk Synchronous Parallel (BSP) model, which was developed to address the problem of parallelizing jobs across multiple workers for scalability 
 The fundamental computing paradigm Pregel employs can be characterized as ``think like a vertex''. 
 Graph computations are specified in terms of what each vertex has to compute; edges are communication channels for transmitting computation results from one vertex to another, and do not participate in the computation.
 To avoid communication overheads, Pregel preserves data locality by ensuring computation is performed on locally stored data. 
 The input graph is loaded once at the start of a program and all computations are executed in-memory. 
 As a result, Pregel supports only graphs that fit in memory \cite{91083}.

Giraph \cite{90505} is an open source implementation of Pregel.
Giraph runs workers as map-only jobs on Hadoop and uses HDFS for data input and output. 
Giraph also uses Apache ZooKeeper for coordination, checkpointing, and failure recovery schemes.
Giraph has incorporated several optimizations, has a rapidly growing user base, and has been scaled by Facebook to graphs with a trillion edges.
Giraph is executed in-memory, which can speed-up job execution, but, for large amounts of messages or big datasets, can also lead to crashes due to lack of memory.

 GraphLab \cite{91044} is an open-source, graph-specific distributed computation platform implemented in C++. 
 GraphLab uses the GAS decomposition (Gather, Apply, Scatter), which is similar to, but fundamentally different from, the BSP model. In the GAS model, a vertex accumulates information about its neighborhood in the Gather phase, applies the accumulated value in the Apply phase, and updates its adjacent vertices and edges and activates its neighbouring vertices in the Scatter phase.
 Another key difference is that GraphLab partitions graphs using vertex cuts rather than edge cuts. Consequently, each edge is assigned to a unique machine, while vertices are replicated in the caches of remote machines.
 Besides graph processing, it also supports various machine learning algorithms.  

There are several works comparing graph processing frameworks.
For instance, the first evaluation study of modern big data frameworks, including Map-Reduce, Stratosphere, Hama, Giraph and Graphlab, is presented in \cite{91089}.
 In \cite{91093}, a benchmarking suite for graph-processing platforms is presented. The suite was used to evaluate the performance of Hadoop, YARN, Stratosphere, Giraph, GraphLab, and Neo4j.
 In \cite{91090}, the authors present a comparison study on parallel processing systems, including Giraph, GPS and GraphLab.
 Finally, an Experimental Comparison of Pregel-like Graph Processing Systems is presented in \cite{91083}.

\subsection{RDF database systems}

An RDF database (also called \emph{Triple Store}) is a specialized graph database for managing RDF data.
RDF defines a data model based on expressions of the form subject-predicate-object (SPO) called RDF triples.
Therefore, an RDF dataset is composed by a large collection of RDF triples which implicitly form a graph.

SPARQL is the standard query language for RDF databases.
It is a declarative language which allows to express several types of graph patterns. Its most recent version (SPARQL 1.1) supports advanced features like property paths, aggregate functions and subqueries.

There are several works comparing RDF databases (see for example \cite{91017}).
Similar to graph databases, RDF databases can also be classified into native and non-native RDF databases.
Examples of native RDF databases are Jena \cite{10053}, RDF-3X \cite{90769}, 4store \cite{91145} and TripleBit \cite{91117}.
Among the non-native RDF databases we can mention to 
OpenLink Virtuoso \cite{90060}, Sesame \cite{90332} and DB2RDF \cite{90801}, which are implemented on top of relational database systems. 

 Being more specific about native storage approaches, the RDF databases can be classified into four categories \cite{91117}: triples table, property table, column store with vertical partitioning and RDF graph based store.

A \emph{triple table} refers to the approach of storing RDF data in a 3-column table with each row representing a SPO statement. 
Hence, the evaluation of SPARQL queries involve self-joins over this long table. 
A popular approach to improving performance of queries in this storage model is to use an exhaustive indexing method that creates a full set of SPO permutations of indexes.
Among the systems implementing triple tables we can mention 
RDF-3X \cite{90769},
Sesame \cite{90332}, 
3store \cite{90331}, 
BrightstarDB \cite{91096}.

A second approach is to store RDF data in a \emph{property table} \cite{90759} with subject as the first column and the list of distinct predicates as the remaining columns.
A single property table can be extremely sparse and contains many NULL values. Thus multiple-property tables with different clusters of properties are  proposed as an optimization technique.
Jena \cite{10053}, Oracle \cite{91143},  and BitMat are examples of systems implementing property tables.

RDF data can also be stored by using multiple two-column tables, one for each unique predicate. The first column is for subject whereas the other column is for object. This method, called \emph{column store with vertical partitioning} \cite{90759}, can be implemented over row-oriented or column-oriented database systems.
This column-store approach is implemented by C-Store

Finally, a graph based approach focuses on storing RDF data as a graph.
In this case, the RDF triples must be modeled as classical graph nodes and edges, and the SPARQL queries must be transformed into graph queries. 
Among the \emph{RDF graph based stores} we can mention
Ontotext GraphDB \cite{91106},
gStore \cite{91133}, 
Stardog \cite{91116}, 
Blazegraph \cite{91075}, 
TrinityRDF \cite{91104} and 
GEMS \cite{91121}.

\bibliographystyle{abbrv}


\end{document}